\RequirePackage{fix-cm}
\documentclass{svjour3}                     
\smartqed  
\usepackage{graphicx}
\usepackage[square,sort,comma,numbers]{natbib}
\usepackage{mathptmx}      
\usepackage{amsmath}
\usepackage{graphicx}
\usepackage{algorithm,algorithmic}
\begin{document}

\title{Approaching unstructured search from function bilateral symmetry detection - A quantum algorithm}

\subtitle{Polynomial time Turing reduction of unstructured search into function bilateral symmetry detection problem}


\author{Dinesh Kumar         \and
        Pankaj Srivastava
}


\institute{Dinesh Kumar \at
           Computational Nanoscience and Technology Lab. (CNTL)\\
              ABV Indian Institute of Information Technology and Management Gwalior, India\\
              Tel.: +91 8950411980\\
              \email{dinesh.gjust@gmail.com}           
           \and
           Pankaj Srivastava \at
           Computational Nanoscience and Technology Lab. (CNTL)\\
ABV Indian Institute of Information Technology and Management Gwalior, India\\
              Tel.: +91-751-2449814\\
              Fax: +91-751-2449814\\
              \email{pankajs@iiitm.ac.in}
}

\date{Received: date / Accepted: date}
\titlerunning{Approaching unstructured search from function bilateral symmetry detection}

\maketitle

\begin{abstract}
Detection of symmetry is vital to problem solving. Most of the problems of computer vision and computer graphics and machine intelligence in general, can be reduced to symmetry detection problem. Unstructured search problem can also be looked upon from symmetry detection point of view. Unstructured search can be thought as searching a binary string satisfying some search condition in an unsorted list of binary strings. In this paper unstructured search problem is reduced to function bilateral symmetry detection problem with polynomial overhead in terms of the size of the input.
\end{abstract}

\section{Introduction}
\label{Sec:intro}
Searching something has always been an interesting computational problem in algorithm design. It's the most primitive and most frequent operation performed in day to day computational tasks. As information is encoded most often in binary strings, searching often means searching a string item in a search space of string items. There have been algorithms designed which work fine for a limited unstructured search space. For faster search operation it is required to maintain some structure in the the search space. The structure may be the order of the elements (sorted list) or some other structure like Search tree, Heap etc. Any way, maintaining the structure is a costly operation as inserting a new element to the existing structure or deleting an element from the existing structure are both costly. So depending upon which operation we want to be perform fast some structure is maintained or no structure is maintained. Sometimes a trade-off is made between the two.\\
Power of quantum computing has been widely established. There are computational tasks which can be performed faster using a quantum computer for which no efficient classical algorithm exists \cite{doi:10.1137/S0097539796298637}, \cite{bernstein1993quantum},  \cite{shor1997polynomial}. In \cite{Grover:1996:FQM:237814.237866} a quantum search algorithm was designed which can solve the unstructured search problem in $O(\sqrt{N})$ time. $N$ is the number of string items in the search space. The Grover's algorithm has found extensive application in solving other computational problems. Quantum algorithms like quantum counting, amplitude amplification and applications like collision problem, finding the median, minimum etc, graph problems (spanning trees, matchings, flows....) and many others are based on Grover's Unstructured search algorithm. In his remarkable work \cite{shor1997polynomial} Peter W. Shor  solved  the problem of prime factorization in polynomial time. Problem of prime factorization can be viewed as unstructured search problem. Prime factorization can be thought as searching the prime numbers in the solution space of integers which divides the composite number whose factors has to be found. Once the solution is known it is easy to verify its correctness. The search part is crucial in solving to most of the problems and time consuming classically.
Performing search on sorted list of items or database is easy and already solved in \cite{grover1997quantum} with just one query complexity. \\
Over the past years it's widely believed that the Grover's unstructured search algorithm is optimized and the time complexity can not be reduced beyond $O(\sqrt{N})$. Many work already exist that shows that the quantum lower bound is exactly optimal \cite{ambainis2000quantum}, \cite{Beals:2001:QLB:502090.502097} and more. Using the quantum random walk model \cite{PhysRevA.67.052307} solved the same search problem in $O(\sqrt{N})$ time. In this paper the unstructured search problem is reduced to function bilateral symmetry detection problem with polynomial overhead in terms of size of input search space in bits. \\
\paragraph{Unstructured Search Problem}
Given a list $L$ of string items of length $m$ bits, find the index $i$ of the string item $s$ in the list that satisfies some search criteria $f_1:{\{0,1\}}^n \rightarrow \{0,1\}$ defined below, if it exists.
\[f_1(i) = \left\lbrace\begin{array}{cc}
1 & \quad \text{if $f_2(L[i])=z$}\\
0 & \quad \text{otherwise}
\end{array} \right. \]
Here $f_2:{\{0,1\}}^m \rightarrow {\{0,1\}}^l$, $m \geq l$, can be any function which runs in polynomial time. $z$ is a bit string of arbitrary length $\leq$ $m$. Setting $f_2$ and $z$ can be thought as specifying the search criteria $f_1$ over $L$. Parameters $L$, $f_2$ and $z$ are fixed for a particular search. Only $i$ is varied over whole search domain. The above search criteria can be read as find the string in the list which results in $z$ when applied to $f_2$. A simple case is when $f_2$ is identity function i.e. when the search condition is to find just the match for $z$ in the list $L$.
It is possible that multiple item strings in the list satisfies the search criteria $f_1$.
So the search problem is to find the indexes for which $f_1(i)$ equals 1.\\
Symmetry detection and its significance in computation has already been recognized \cite{hel2010computational}.
Mathematically, a symmetry can be thought as a mapping or transformation from a set of points $S$ of an object into itself such that the whole object remains invariant after the transformation. The concept of symmetry has found its application in almost all fields of study including but not limited to Arts, Mathematics, Physics, Biology, Chemistry and  Social interaction. Symmetry is inherent in nature, so in sciences. \\
In the following section, concept of function bilateral symmetry is discussed. Next, conversion of unstructured search problem into symmetry detection problem is presented. Finally two quantum algorithms are proposed to solve unstructured search problem. Both algorithms use the concept of function bilateral symmetry detection for solving the search problem.

\section{Function bilateral symmetry detection} 
\label{Sec:FunBilSymDet}
\begin{figure}[!h]
 \centering
  \mbox{\includegraphics[width = .60\textwidth]{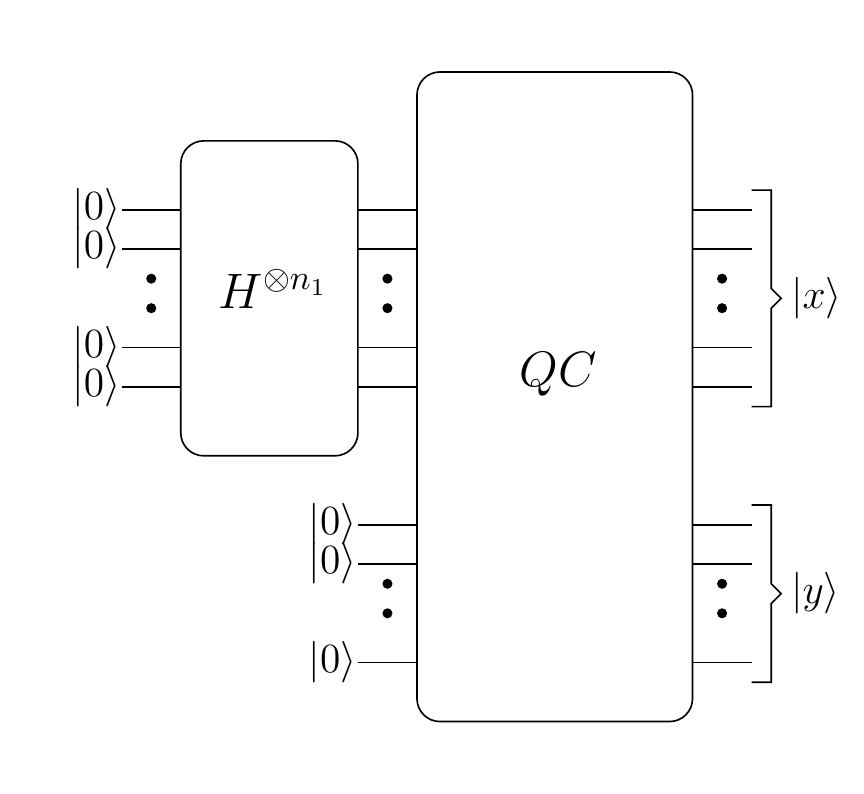} }
    \caption{Function bilateral symmetry [$EvenOrNot()$] detection algorithm }
    \label{Fig:EvenOrNot}
  \end{figure}

A function $f$ possesses bilateral symmetry or reflection symmetry if it is even i.e. $f(x) = f(-x)$. This section discusses the concept of function bilateral symmetry detection.\\~\\
\textbf{Problem Definition}:\\
Given a black box function $f$ from ${\{0,1\}}^{n_1}$ to ${\{0,1\}}^{m_1}$, determine weather the function is symmetrical about y axis or not.\\~\\ 
In other words we have to check weather the function is a even function or not. Function bilateral symmetry detection is a decision problem. Classically, we have to query (call) the black box function for all its input domain and make one halve the comparison of the number of elements in the input domain to determine the evenness of the function $f$. That is weather $f(x) = f(-x)$ $\forall$ $x$ $\epsilon$ [$+ve$ $domain$ of $f$]. It requires $O(2^{n_1})$ steps to do so. Let us assume that we can design a quantum algorithm for function bilateral symmetry detection with time overhead of $O(q = F(n_1))$, where $F$ is some yet to be known function of $n_1$. 
Figure \ref{Fig:EvenOrNot} refers to the the high level design of such a quantum algorithm.
 

QC is some quantum circuit which makes use of $f$. $y$ is the output register. The input register is initialized to $|000..00\rangle$ state. The circuit outputs $y = 00..0$ if $f$ possesses bilateral symmetry. A non zero value of $y$ indicates that the function $f$ is not an even function. Above algorithm can be referred to as $EvenOrNot(f,i)$, where $f$ is the function whose symmetry needs to be tested and $i$ refers to the input register of $EvenOrNot$ detection algorithm. Value of register $i$ can be used to control the input domain for which evenness of $f$ needs to be tested. For example a value $|+\rangle|-\rangle|0\rangle|0\rangle....|0\rangle$ of register $i$ can be used to test the evenness of $f$ for all possible input values starting with $01$.
\section{Reducing unstructured search into function bilateral symmetry detection problem} 
\begin{figure}[!h]
\centering
\includegraphics[scale=.50]{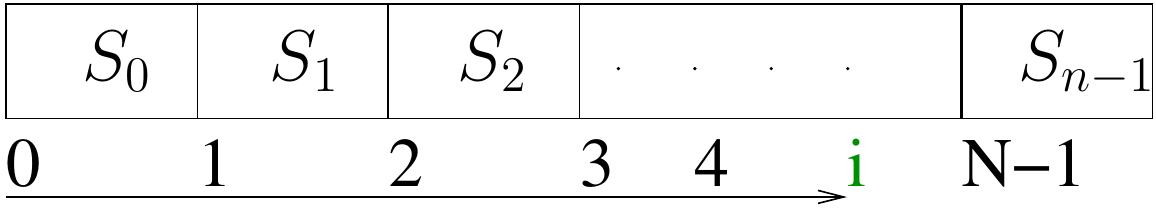}
\caption{Unstructured search problem}
\label{fig:UnsortedList}
\end{figure}

\begin{figure}[!h]
\centering
\includegraphics[scale=.50]{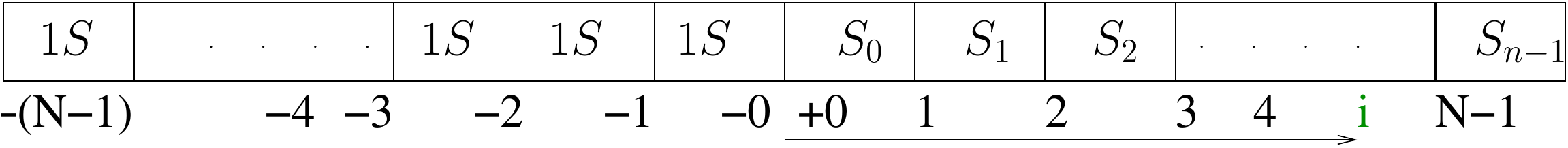}
\caption{New search problem}
\label{fig:Modified List}
\end{figure}

Function bilateral symmetry detection is a decision problem as the answer to the problem is yes or no only. Unstructured search requires finding the index $i$ of the searched item $s$ in the list which satisfy some search condition $f_1$. Index $i$ is $log_2N$ bit string where $N$ is the size of the search space. Determining $i$ requires making decisions about individual bits of $i$. The decision weather the bit value should be 0 or 1 has to be made for all $log_2N$ bits. In this way unstructured search can be viewed as a decision problem.
Conversion of the search problem into the symmetry detection problem requires the search problem to be converted to some function such that its symmetry checking helps in determining the bits of index $i$. If we query (call) the function $f_1()$ one by one for all the items in the list $L$ (fig. \ref{fig:UnsortedList}) then at some index $i$ we get 1 as output and remaining of the times we get 0 as result of the query to $f_1()$. To check the symmetry, we need to define function $f_1()$ for negative values too. Let's define a new function $f_3:{\{0,1\}}^{n+1} \rightarrow \{0,1\} $ as follows:
\[f_3(i) = \left\lbrace\begin{array}{cc}
f_1(i_{n-1}i_{n-2}...i_{1}i_{0}) & \quad \text{if $i_n = 0$}\\
0 & \quad \text{otherwise}
\end{array} \right. \]
\hspace{-1ex} The newly defined search criteria $f_3$ can be thought as searching the same string item $s$ in the big list (figure \ref{fig:Modified List}) of length $2N$ string items.
\hspace{-1ex}Signed number representation is used for index $i$ values. Due to this there are two representation of 0 in the new list (00..00 is +ve zero, 10..00 indicates -ve zero). The MSB determines the sign of the number and rest bits determine the magnitude of the number. The prepended string $1S$ in the new list can be any arbitrary string of length $m+1$ bits. As the solution index $i$ can not be -ve as per definition of $f_3$, it will remain the same for the new search problem except now it will be a $n+1$ bit number with MSB 0 (positive number). The newly defined function $f_3()$ is checked for bilateral symmetry (evenness). A non zero output from $EvenOrNot()$ quantum algorithm shows that string item $s$ satisfying the search criteria $f_3()$ is present in the list. Next, We divide the positive domain of $f_3()$ in two equal halves. Check the evenness of $f_3()$ for both positive halves with corresponding negative halves. The item $s$ must be present in the half in which the function is not even. We divide that half further and repeat the above procedure until the domain of $f_3()$ is reduced to contain only a single integer index $i$.
 
\section{Quantum unstructured search algorithm} 
\subsection{For single item search}
Algorithm \ref{Algo:Quantum Unstructured Search Algo} describes the pseudo code. It takes as input the search criteria $f_3()$ and a $(n+1)$ qubit register $i$ as input. Search criteria $f_3()$ is constructed by setting $f_1$ which in turn requires setting $f_2$, $L$ and $z$. At any point of time during the algorithm execution, $i$ is indirectly used to refer to the domain of the item strings in which searched string $s$ satisfying the search criteria $f_3$ could be present. In fact, applying Hadamard transform $H^{\otimes (n+1)}$ to $i$ results in a superposition state which corresponds to the domain of item   
\begin{algorithm}[H]
\caption{Quantum Unstructured Search Algo to find single searched item}
\label{Algo:Quantum Unstructured Search Algo}
\begin{algorithmic}
\STATE{\textbf{Algo QuantumUnstructuredSearch}($f_3(),i$)
 \STATE{Initialize: Register i[ ] $\leftarrow$ \{$|0\rangle,|0\rangle,|0\rangle$ ... upto $log_2N$\}} 
  \STATE{$i[0]$ $\leftarrow$ $|+\rangle$}
 \IF{$EvenOrNot(f_3,i)$ = 0}
          \PRINT "Item not present" 
          \STATE{$exit()$}
          
  \ENDIF        
  \FOR{$j$ $\leftarrow$ 1 \TO $log_2N$}
\STATE  {$i[j]$ $\leftarrow$ $|+\rangle$
\IF{$EvenOrNot(f_3,i)$ = 0}
       \STATE{$i[j]$ $\leftarrow$ $|+\rangle$}
 
      \ELSE
       \STATE{$i[j]$ $\leftarrow$ $|-\rangle$}
	  \ENDIF
  } 
  \ENDFOR
  \STATE{Apply $H^{\otimes (n+1)}$ on register $i$}
  \PRINT "String item present at location $i$"
}
\end{algorithmic}
\end{algorithm}

\hspace{-3ex}strings in which searched string $s$ could be present. For example its value $|+\rangle|+\rangle|-\rangle|0\rangle|0\rangle$ $ ....|0\rangle$ indicates that index of searched string $s$ lies in the set of all $(n+1)$ bit values starting with 001. At the end of the algorithm it will refer to the particular index $i$ of the searched item $s$. The first $if$ statement in the algorithm checks weather the searched item $s$ is present in the list or not. $EvenOrNot(f_3(),i)$ will return $0$ only if $f_3$ is even otherwise it will return a non-zero value. If $EvenOrNot(f_3(),i)$ returns $y$ equal to zero it indicates that the searched item is not present in the list. In case searched item is present in the search domain of $f_3$ then atleast for one value in the search domain, $f_3$ will be 1. It will result into uneven detection and $EvenOrNot()$ algorithm will return $y$ equal to some non-zero value. If the searched item is present in the list then the positive domain of $f_3()$ is divided into two equal halves. Evenness is checked for $i$ equal to $|+\rangle|+\rangle|0\rangle0\rangle..... |0\rangle$ and $|+\rangle|-\rangle|0\rangle0\rangle.....|0\rangle$. The former value of $i$ corresponds to left sub-domain and later value of $i$ corresponds to right sub-domain of the divided domain. Depending upon the result, $i$ is set. If the searched item is present in the left sub-domain, corresponding qubit of $i$ distinguishing two sub-domains is set to $|+\rangle$. Otherwise it is set to $|-\rangle$. The above procedure is repeated until we determine all qubits of $i$. Finally a Hadamard Transform is performed on register $i$ to get the index of the searched item $s$.     
\begin{algorithm}[H]
\caption{Quantum Unstructured Search Algo to find one or more items}
\label{Algo:Quantum Unstructured Search Algo2}
\begin{algorithmic}
\STATE{\textbf{Algo QuantumUnstructuredSearch}($f_3(x)$, $i$, $k$)
  \STATE{$i$[0] $\leftarrow$ $|+\rangle$}
 \IF{k = 1 \AND $EvenOrNot(f_3,i)$ = 0}
         \PRINT "Item not present" 
          \STATE{$exit()$}
          
  \ENDIF
    \FOR{$j$ $\leftarrow$ $k$ \TO $log_2N$}
    \STATE{
  \STATE{$i[j]$ $\leftarrow$ $|+\rangle$}        
  \STATE{$y_1$ $\leftarrow$ $EvenOrNot$($f_3$,$i$)}
  \STATE{$i[j]$ $\leftarrow$ $|-\rangle$}
  \STATE{$y_2$ $\leftarrow$ $EvenOrNot$($f_3$,$i$)}
  \IF{$y_1$ $\neq$ 0 \AND $y_2$ = 0}
   \STATE{$i[j]$ = $|+\rangle$}
  \ELSE
  {
    \IF{$y_1$ = $0$ \AND $y_2$ $\neq$ $0$}
    \STATE{$i[j]$ $\leftarrow$ $|-\rangle$}
    \ELSE
	  \STATE{$i[j] \leftarrow$ $|+\rangle$ \\ 
	  $QuantumUnstructuredSearch$($f_3$, $i$, $k+1$) \\
	    $i[j]$ $\leftarrow$ $|-\rangle$  \\
	  $QuantumUnstructuredSearch$($f_3$, $i$, $k+1$) \\
	  $exit()$
	   }
	  \ENDIF
  }
  \ENDIF  
  }
  \STATE{$k$ = $k$ + 1}
  \ENDFOR
    \STATE{Apply $H^{\otimes (n+1)}$ on register $i$}
    \PRINT "String item present at location $i$"
  }
\end{algorithmic}
\end{algorithm}

\subsection{For multiple item search}
Algorithm \ref{Algo:Quantum Unstructured Search Algo2}  defines the pseudo code for multiple string item search. The algorithm works in a similar way as in the case for single item search. Search condition $f_3(x)$ is constructed by setting $f_1$ which in turn requires setting $s$, $f_2$ and $L$. Register $i$ is passed with initial value $i$[ ] = \{$|0\rangle,|0\rangle,|0\rangle$ ... upto $log_2N$\}. Register $k$ stores the bit position of index $i$ such that all bits of $i$ to the left of bit position $k$ have been determined. It is passed with initial value $k = 1$ as the first qubit of the solution index $i$ is bound to be 0 if an item is present in the list. The first if statement of the algorithm checks weather any searched item is present or not. The following for loop runs $log_2N$ times. Each iteration of the for loop determines one qubit of the solution indexes. In each iteration the domain is divided into two halves. There can be three cases possible for the searched strings. All the  searched strings can be present in the left sub-domain, right sub-domain or both sub-domains simultaneously. In case when the items are present in both sub-domains the algorithm calls itself recursively for both sub-domains. The newly called instances of the algorithm starts their work from the smaller sub-domain for which they are called.
\section*{Complexity analysis}
\paragraph{Search for single item:}For a search space of $N$ items the algorithm requires $(log_2N+1)$ calls to $EvenOrNot()$ quantum algorithm. First call to $EvenOrNot()$ determines the presence of the item in the list and the rest $log_2N$ calls determines a bit of solution index $i$. If $EvenOrNot()$ quantum algorithm takes $O(q = F(n_1))$ time, the total time complexity of the algorithm for single item search will be $O(qlog_2N)$.
\paragraph{When two items are present:} Let $i_1$ and $i_2$ are  $n$ bit indexes of the searched items present in the list meeting the search criteria. Let $i_1$ and $i_2$ be same for $a$ MSB's (Most significant bits) i.e. $\underbrace{b_{n-1}b_{n-2}b_{n-3}...b_{n-a}}_a\underbrace{b_{n-a-1}b_{n-a-2}....b_{1}b_{0}}_{n-a}$. The algorithm will make $(a+1)$ calls to $EvenOrNot()$ to determine the first $a$ bits of $i$. For the remaining $(n-a)$ bits, two separate instances of algorithm \ref{Algo:Quantum Unstructured Search Algo2} will be running. Each instance will make $(n-a)$ calls to $EvenOrNot()$ algorithm. Hence total calls will be $((a+1) + 2(n-a)) \Rightarrow O(n)$ $\Rightarrow$ $O(log_2N)$ and time complexity for this case will be $O(qlog_2N)$.
\paragraph{In case of more than two items:}Let $i_1$, $i_2$...$i_c$ be the indexes of the string items matching the search criteria. Let $a$ initial MSB's are same for all the indexes. Then a total $(a+1)$ steps will be required to determine initial $a$ bits of all indexes. The indexes can be divided into two different classes or sets depending upon the similarity in initial ${(a+1)}$ bits. The above procedure of division of indexes is continued until all bits of all indexes are determined. In a loose setting, searching $c$ string items can be thought as running the search algorithm \ref{Algo:Quantum Unstructured Search Algo} $c$ times finding a new index in each search. So The upper bound is $clog_2N$ times. In this way, it still requires $O(qlog_2N)$ steps to solve the problem.

\section{Conclusion}
\label{sec:con}
Looking at the unstructured search problem from the view point of symmetry is a completely different approach towards solving the search problem. Even though no efficient algorithm to solve the function bilateral symmetry detection problem is known yet, function bilateral symmetry detection may have more structure to it that may be utilized to solve the search problem more efficiently. Solving unstructured search problem can be viewed as making a sequence of decisions. This paper establishes that function bilateral symmetry detection problem is atleast as hard as solving the unstructured search problem.
Most of the problems can be viewed as search problem. Problem of prime factorization and 3-SAT are examples of such problems. Such problems can be viewed as unstructured search problem with the verification function $f_2$ (problem dependent) as part of the search criteria $f_1$. Verification function $f_2$ tests weather a given solution is correct or not. If an efficient quantum algorithm can be designed for the function bilateral symmetry detection, we can solve the unstructured search problem and many other problems efficiently using quantum computing.


%


\bibliographystyle{spbasic}      
\bibliography{ApproachingUnstructuredSearchDifferently}


\end{document}